\documentclass[final]{aipproc}

\layoutstyle{6x9}
\begin{document}
\newcommand{\stp}{\tilde t_1}
\def\chio{\textstyle\raise.4ex\hbox{$\textstyle\tilde\chi{^0_1}$}}

\begin{titlepage}

\thispagestyle{empty}
\def\thefootnote{\fnsymbol{footnote}}       

\begin{center}
\mbox{ }

\end{center}
\begin{flushright}
\vspace* {-2.0cm}
\Large
\mbox{\hspace{10.2cm} hep-ph/0611351} \\
\end{flushright}
\begin{center}
\vskip 1.0cm
{\Huge\bf
Small Visible Energy Scalar Top
}
\vspace{2mm}

{\Huge\bf
Iterative Discriminant Analysis
}
\vspace{2mm}

{\Huge\bf
for Different Center-of-Mass Energies
}
\vskip 1cm
{\LARGE\bf A. Sopczak$^1$, A. Finch$^1$, A. Freitas$^2$, \\
          C. Milst\'ene$^3$, M.~Schmitt$^4$
\smallskip

\Large $^1$Lancaster University, UK; $^2$Zurich University, Switzerland; \\
       $^3$Fermilab, USA; $^4$Northwestern University, USA}

\vskip 2.5cm
\centerline{\Large \bf Abstract}
\end{center}

\vskip 2.cm
\hspace*{-0.5cm}
\begin{picture}(0.001,0.001)(0,0)
\put(,0){
\begin{minipage}{\textwidth}
\Large
\renewcommand{\baselinestretch} {1.2}
Light scalar top quarks with a small mass difference with respect to the neutralino mass are
of particular cosmological interest. This study uses an Iterative Discriminant Analysis 
method to optimize the expected selection efficiency at a International Linear Collider (ILC). 
A previous study at $\sqrt{s}=260$~GeV with 50~fb$^{-1}$ has been extended to $\sqrt{s}=500$~GeV
with 500~fb$^{-1}$, and results from both studies are compared.
\renewcommand{\baselinestretch} {1.}

\normalsize
\vspace{5.5cm}
\begin{center}
{\sl \large
Presented at SUSY06, the 14th International Conference on Supersymmetry and the Unification of Fundamental Interactions, 
UC Irvine, California, 2006, \\
to be published in the proceedings.
\vspace{-6cm}
}
\end{center}
\end{minipage}
}
\end{picture}
\vfill

\end{titlepage}

\newpage
\thispagestyle{empty}
\mbox{ }
\newpage
\setcounter{page}{1}

\title{Small Visible Energy Scalar Top Iterative Discriminant Analysis for Different Center-of-Mass Energies}


\author{A. Sopczak}{
  address={Lancaster University, UK}
}
\author{A. Finch}{
  address={Lancaster University, UK}
}

\author{A. Freytas}{
  address={Zurich University, Switzerland}
}

\author{C. Milst\'ene}{
  address={Fermilab, USA}
}

\author{M. Schmitt}{
  address={Northwestern University, USA}
}

\begin{abstract}
Light scalar top quarks with a small mass difference with respect to the neutralino mass are
of particular cosmological interest. This study uses an Iterative Discriminant Analysis 
method to optimize the expected selection efficiency at a International Linear Collider (ILC). 
A previous study at $\sqrt{s}=260$~GeV with 50~fb$^{-1}$ has been extended to $\sqrt{s}=500$~GeV
with 500~fb$^{-1}$, and results from both studies are compared.
\vspace*{-2mm}
\end{abstract}

\maketitle

\section{Introduction}
\vspace*{-4mm}
The search for scalar top quarks and the determination of their parameters in the framework of
Supersymmetric models are important aspects of the Linear Collider physics programme.
A light scalar top could play an important role for baryogenesis and, if the mass difference to 
the lightest neutralino is small, for the evolution of the dark matter density through 
co-annihilation~\cite{Balazs:2004bu}. 
This scenario could be explored with precise measurements of the stop properties, in particular its mass, 
which would be possible at the ILC~\cite{lcws05,carena,sopczak}.
This study applies an Iterative Discriminant Analysis (IDA)~\cite{ida} to a scenario 
involving a 122.5~GeV scalar top and a 107.2~GeV neutralino at $\sqrt{s}=500$~GeV.
The study complements the previous study for the same scenario at 
$\sqrt{s}=260$~GeV~\cite{lcws06}.
A luminosity of 500~fb$^{-1}$ is assumed.
The event generation and detector simulation have been performed with unpolarised 
beams as for a sequential cut-based analysis~\cite{carena}. 
In particular, a vertex detector concept of the Linear Collider Flavour 
Identification (LCFI) collaboration~\cite{lcfi}, which studies pixel detectors for 
heavy quark flavour identification, 
has been implemented in simulations for c-quark tagging in scalar top studies.

\vspace*{-4mm}
\section{Preselection}
\vspace*{-4mm}
Only two c-quarks and missing energy (from undetected neutralinos) are expected from\,the reaction 
$\rm e^+e^- \rightarrow \stp \bar{\tilde{t}}_1 \rightarrow c \chio \bar c \chio$.
The requirement $0.1 < E_{\rm vis}/\sqrt{s} < 0.3 $ reduces the
$\rm \rm e^+e^- \to W^+W^-$, ZZ, $\rm q\bar{q}$ and 
$\rm \gamma\gamma \to q\bar{q}$ backgrounds.
Remaining two-photon events are almost completely removed by the cut $p_{\rm t}({\rm event}) > 15$~GeV. 
Requiring at least four but no more than 50 tracks removes mostly very low multiplicity background. Furthermore,
cuts are applied on the longitudinal imbalance $| p_{\rm long}/p | < 0.9$ and thrust angle $| \cos{\theta}_{\rm th}| < 0.95$.
The effect of this preselection is summarized in Table~\ref{tab:pre}.

\clearpage
\begin{table}[h]
\begin{tabular}{l|r|r|r|r|r|r}
        & Total         & 50\%     & After  &$\sigma$            & Factor          & Expected     \\
Process & $\times 1000$ & training & preselection  & (pb)         & per 500~fb$^{-1}$& events       \\ 
\hline
signal  & 50      & 25         & 11115                   & 0.118                & 2.36              &  26232    \\

$\rm q \bar{q}$,  $\rm q \neq t$   
         & 350    & 175      &    14                 & 13.14\phantom{.0}        & 44.5\phantom{0}           & 623     \\  

$\rm W^+W^-$ & 210 & 105       &    11                 & 8.55\phantom{.0}         & 40.7\phantom{0}         &  448     \\

$\rm W e\nu$ & 210 & 105       &    603                &  6.14\phantom{.0}        & 29.2\phantom{0}          & 17633  \\

2-photon & 8750 &  4375        &     22                & 936\phantom{.000}          & 105\phantom{.00}           & 2314   \\        

$\rm ZZ$ & 30   &  15        &      5                &  0.49\phantom{.0}        & 16.8\phantom{0}            & 84     \\

$\rm e e Z$  & 210   & 105     &      4                &  7.51\phantom{.0}        &  37.5\phantom{0}          &   150    \\

$\rm t \bar{t}$ & 180  & 90    &      0                &  0.55\phantom{.0}        &  3.05          &             0    \\

\end{tabular}
\caption{Generated events, events used for the IDA training, events after the 
preselection, $\sqrt{s} = 500$ GeV cross section, scaling factor, and expected number of events.}
\label{tab:pre}
\end{table}

\begin{figure}[htbp]
\begin{minipage}{0.49\textwidth}
\includegraphics[width=\textwidth]{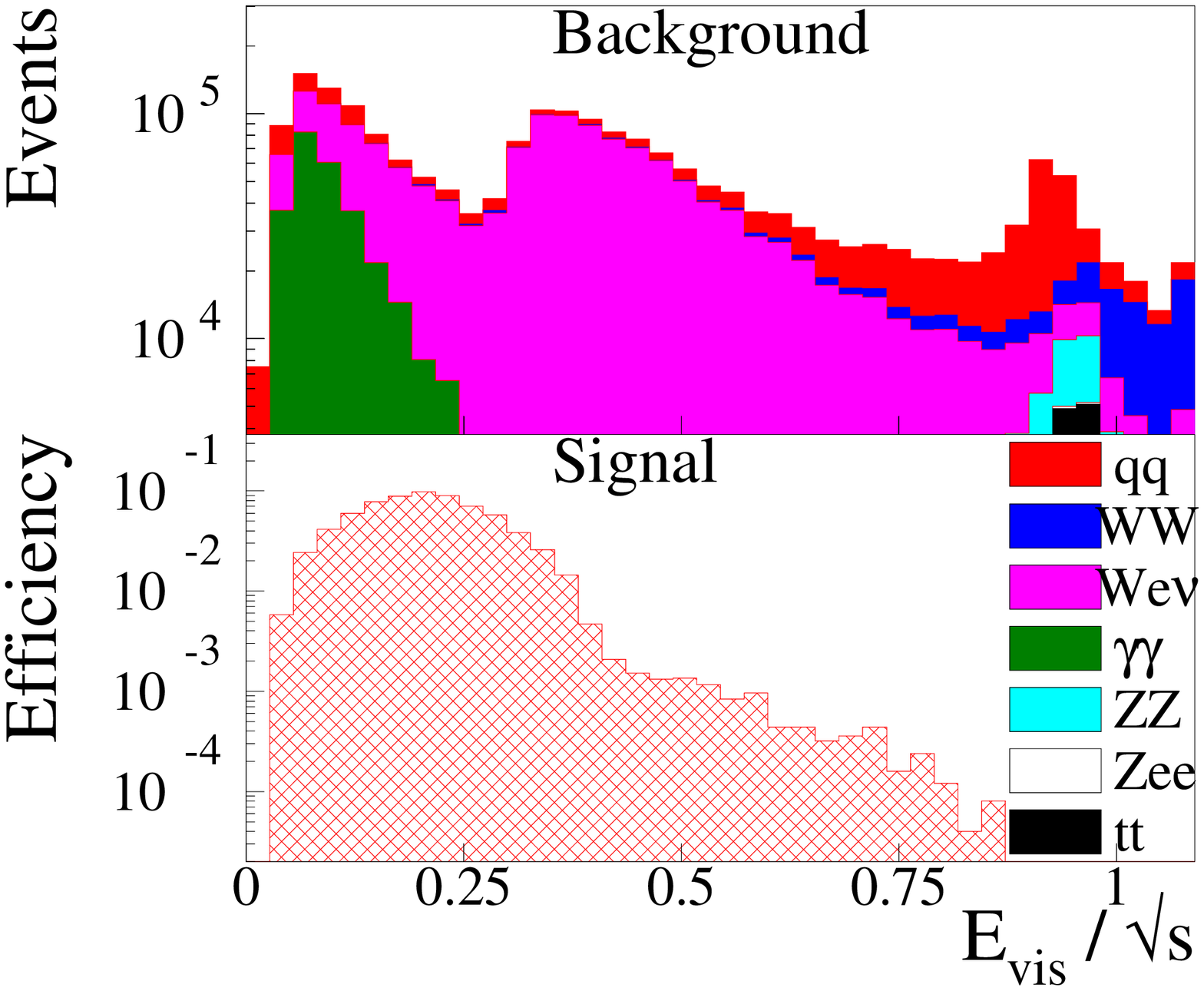}
\end{minipage}
\begin{minipage}{0.49\textwidth}
\includegraphics[width=\textwidth]{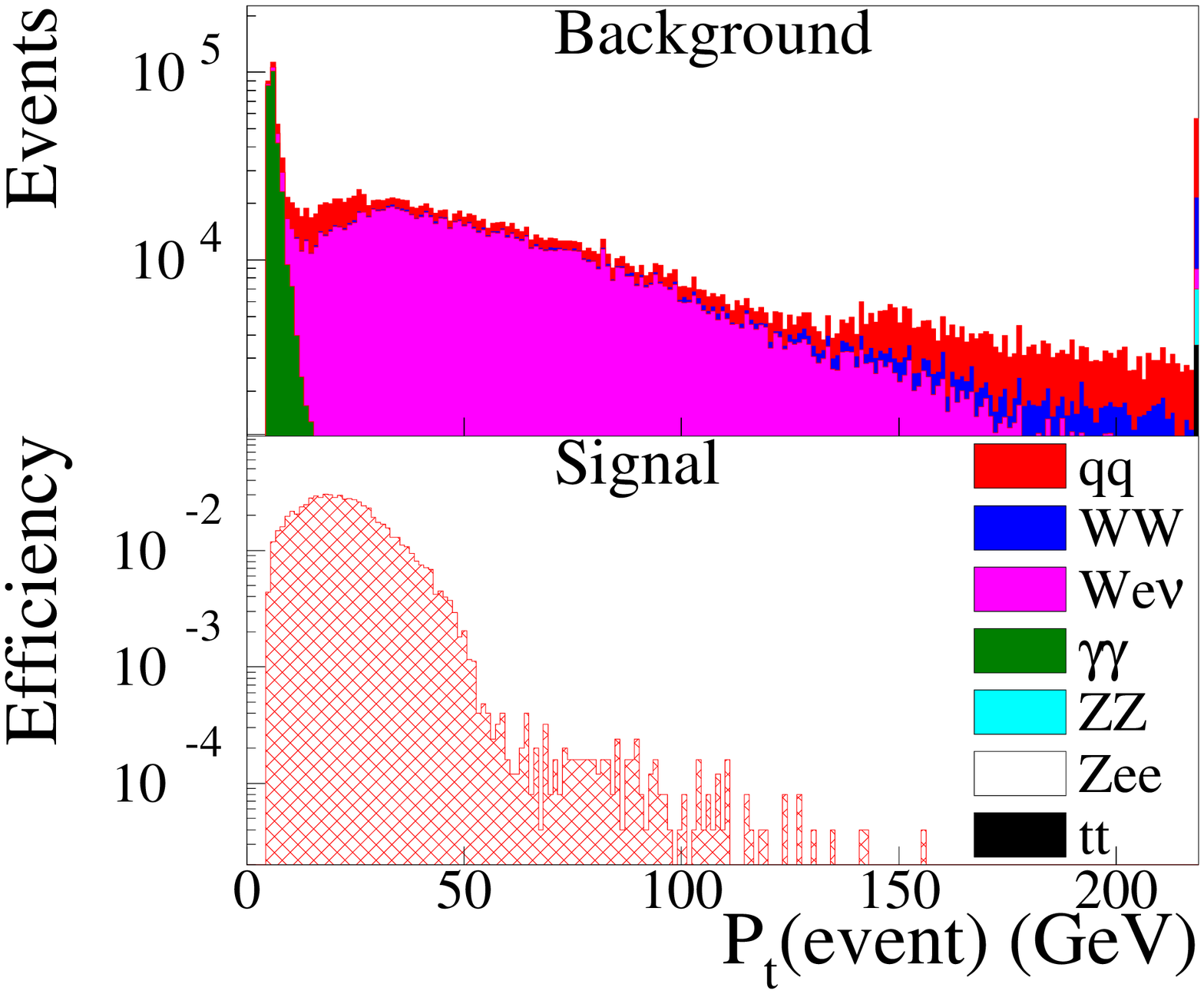}
\end{minipage}
\vspace*{-0.6cm}
\caption{Event distributions before preselection for 500~fb$^{-1}$ at $\sqrt{s} = 500$ GeV.}
\label{fig:pre}
\vspace*{-0.7cm}
\end{figure}

\begin{figure}[H]
\begin{minipage}{0.49\textwidth}
\includegraphics[height=5.5cm,width=\textwidth]{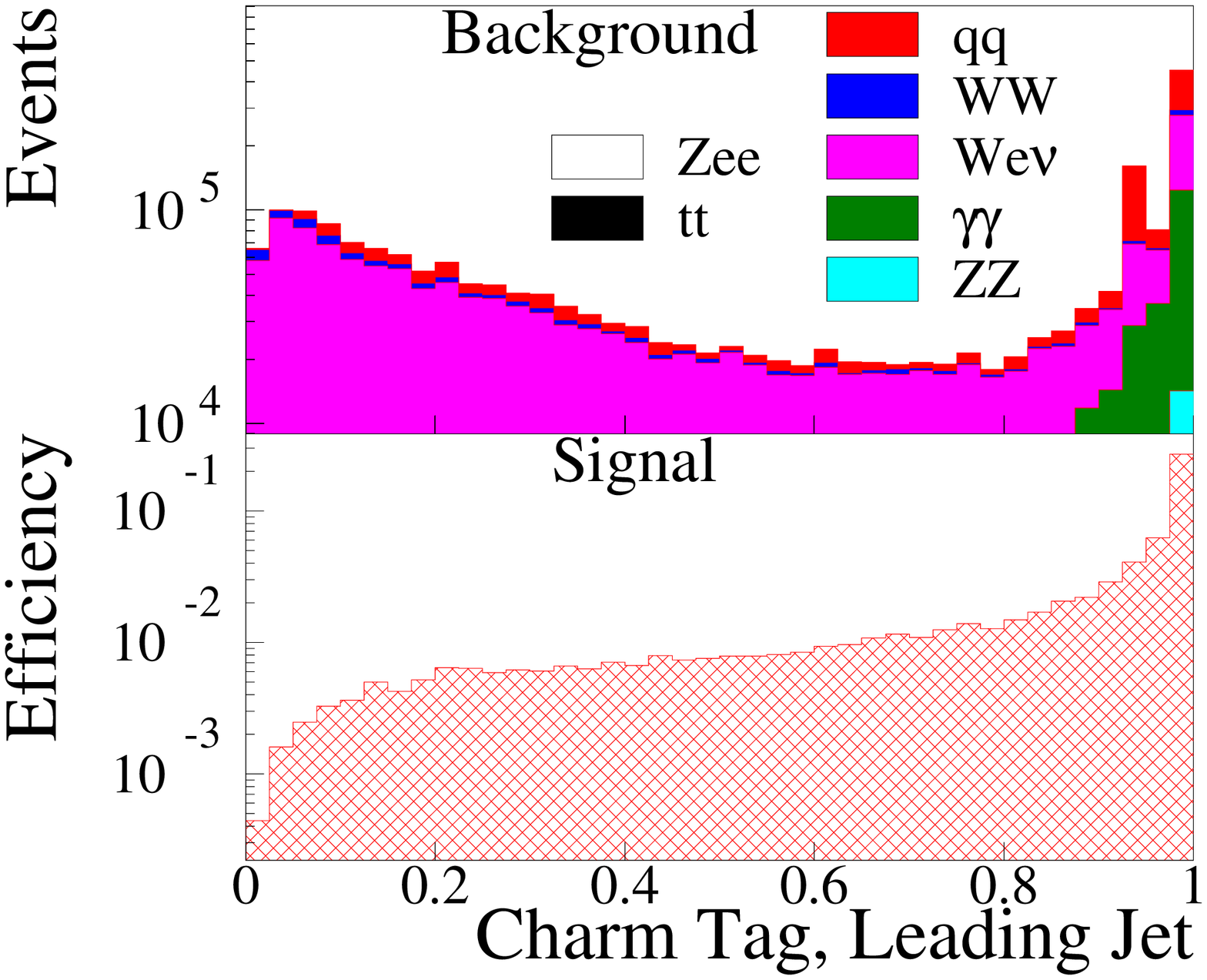}
\end{minipage}
\begin{minipage}{0.49\textwidth}
\includegraphics[height=5.5cm,width=\textwidth]{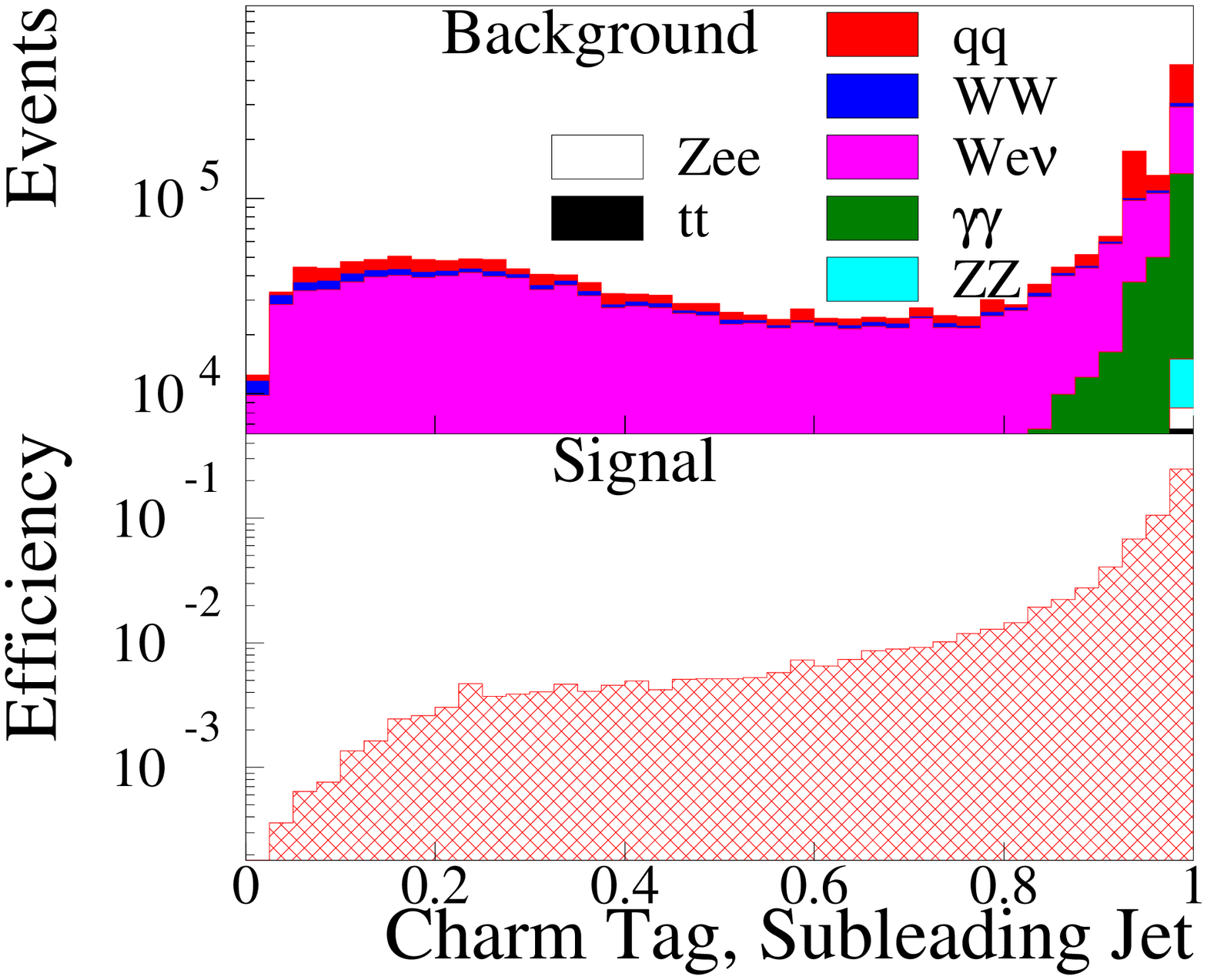}
\end{minipage}
\caption{Distributions of IDA inputs before preselection for 500~fb$^{-1}$ at $\sqrt{s} = 500$ GeV.}
\label{fig:ctag}
\end{figure}

\vspace*{-0.2cm}
The distributions of visible energy and transverse momentum before preselection 
are shown in Fig.~\ref{fig:pre}.
Additional input variables have been used in the IDA:
the event invariant mass and the invariant mass of the two jets. 
Further input variables are the c-quark tag of the leading (most energetic) 
and subleading jets (Fig.~\ref{fig:ctag}). The c-quark jet tagging has been performed
with a neutral network~\cite{kuhl} optimized for small $\Delta m$.

The preselection signal efficiency of 44.5\% at $\sqrt{s} = 500$ GeV 
is slightly lower compared to the 52.5\% for the $\sqrt{s} = 260$ GeV study~\cite{lcws06}.
The corresponding preselection background rates of 21251 events per 500~fb$^{-1}$ 
at $\sqrt{s} = 500$ GeV and 2379 background events per 50~fb$^{-1}$ at $\sqrt{s} = 260$ GeV
are about equal.

\vspace*{-4mm}
\section{IDA Event Selection}
\begin{figure}[H]
\begin{minipage}{0.49\textwidth}
\includegraphics[height=5.5cm,width=\textwidth]{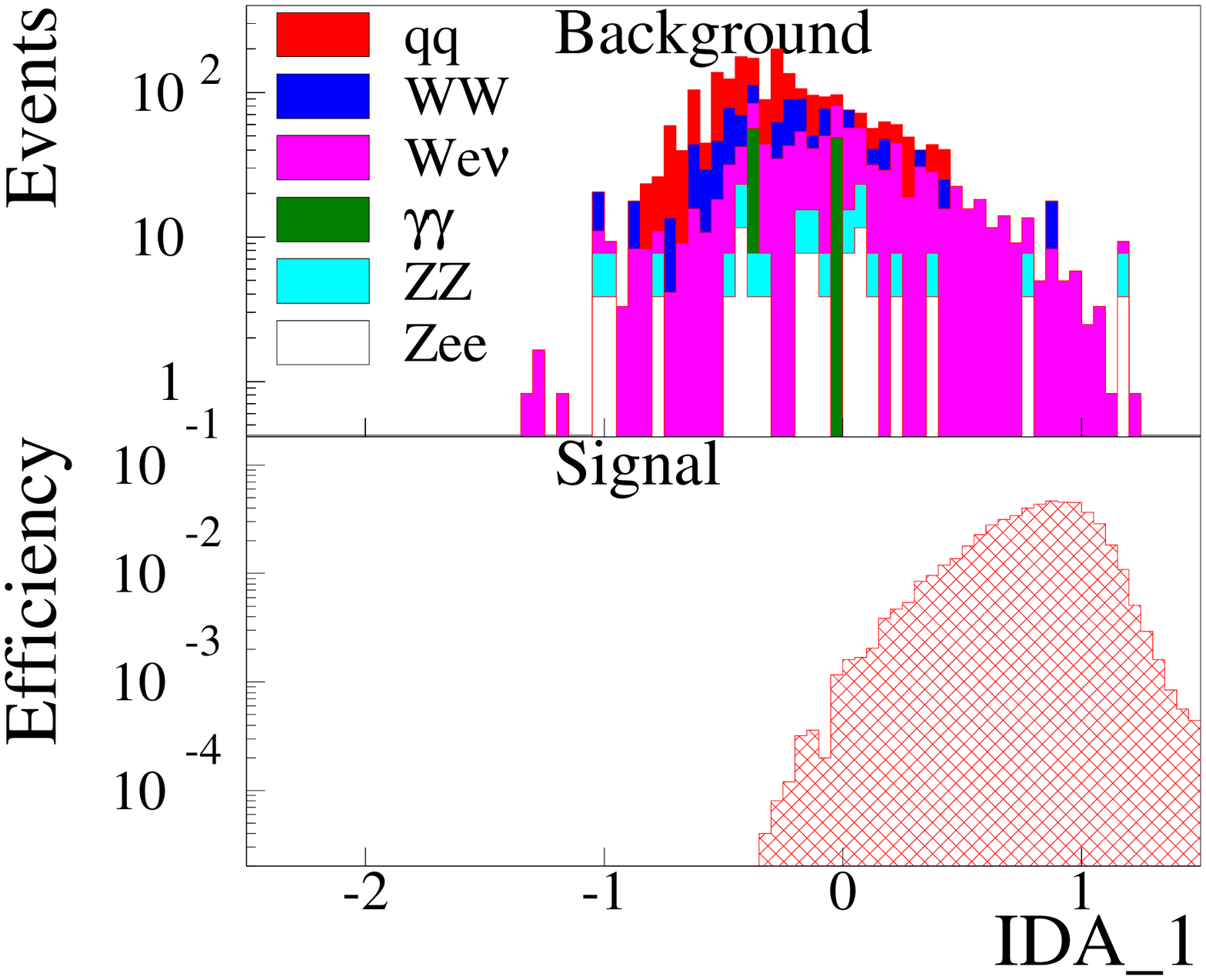}
\end{minipage}
\begin{minipage}{0.49\textwidth}
\includegraphics[height=5.5cm,width=\textwidth]{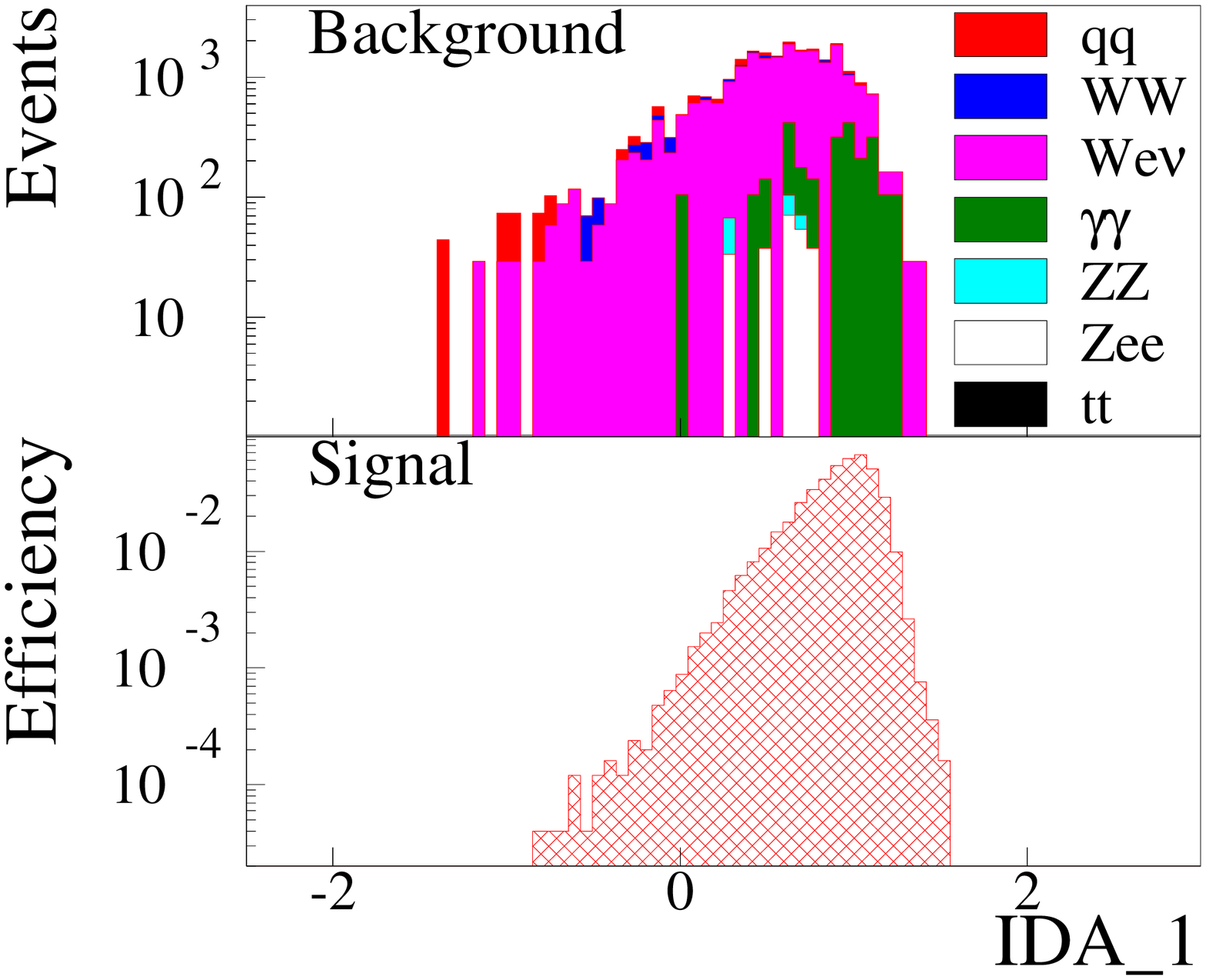}
\end{minipage}
\caption{IDA output for steps 1 for 50~fb$^{-1}$ at $\sqrt{s} = 260$ GeV (left plot) 
         and for 500~fb$^{-1}$ at $\sqrt{s} = 500$ GeV  (right plot).
         In the first IDA step, a cut on IDA\_1 is applied at zero,
         retaining 99.5\% of the simulated signal input events.}
\label{fig:ida}
\end{figure}

\begin{figure}[H]
\begin{minipage}{0.49\textwidth}
\includegraphics[height=5.5cm,width=\textwidth]{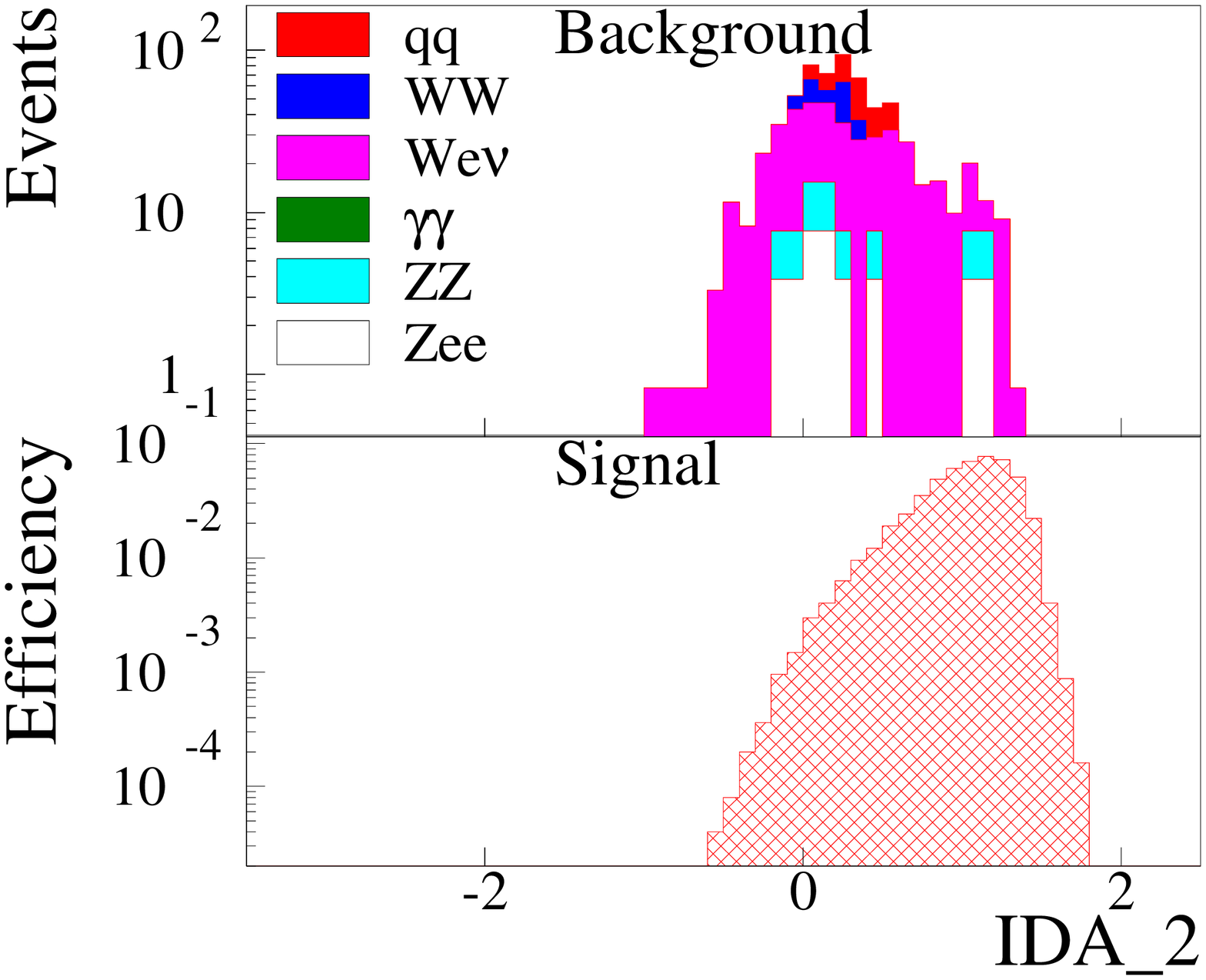}
\end{minipage}
\begin{minipage}{0.49\textwidth}
\includegraphics[height=5.5cm,width=\textwidth]{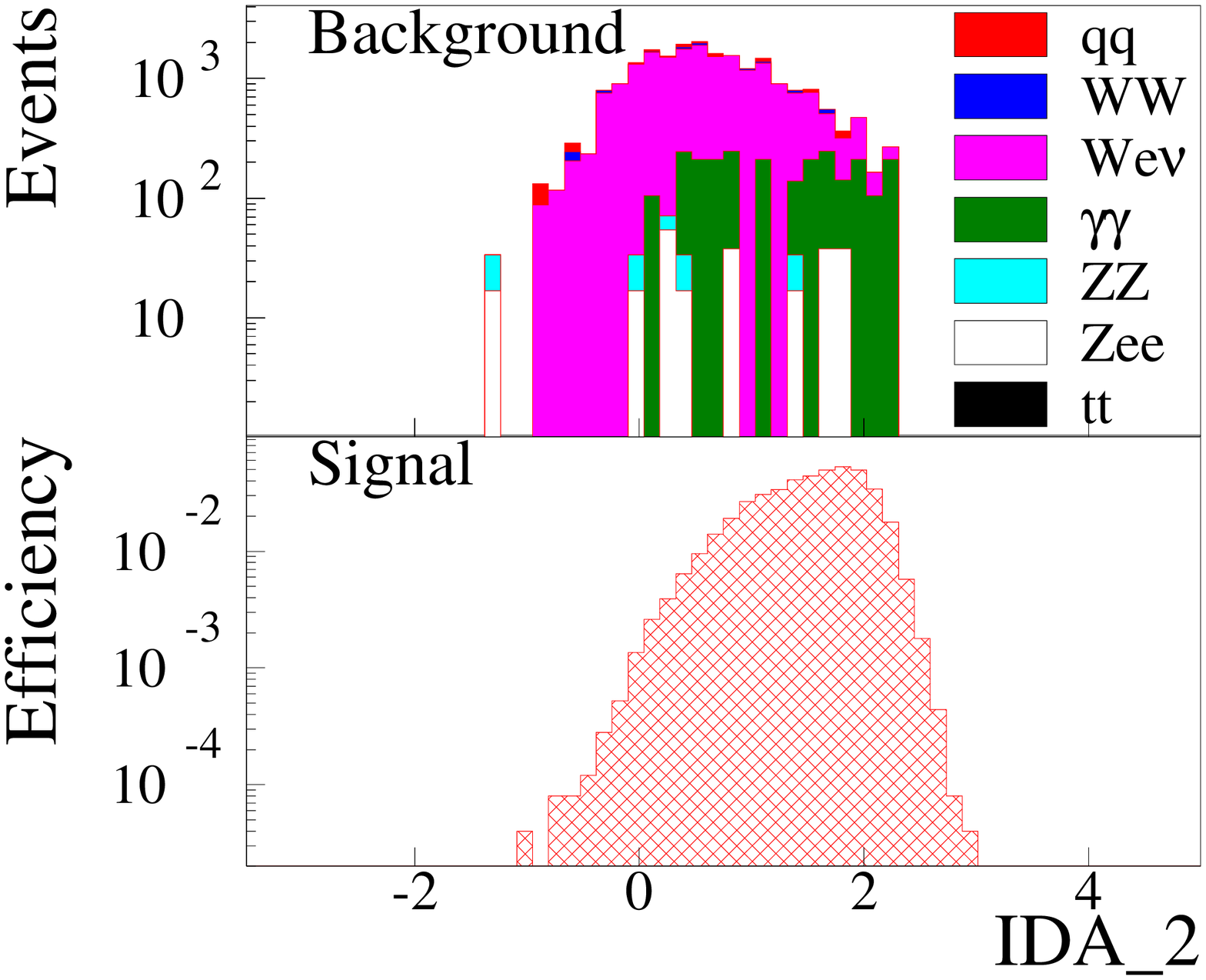}
\end{minipage}
\caption{
IDA output for steps 2 for 50~fb$^{-1}$ at $\sqrt{s} = 260$ GeV (left plot) and 
         for 500~fb$^{-1}$ at $\sqrt{s} = 500$ GeV (right plot).}
\label{fig:ida2}
\end{figure}

\begin{figure}[htbp]
\begin{minipage}{0.49\textwidth}
\vspace*{-1cm}
\includegraphics[width=\textwidth]{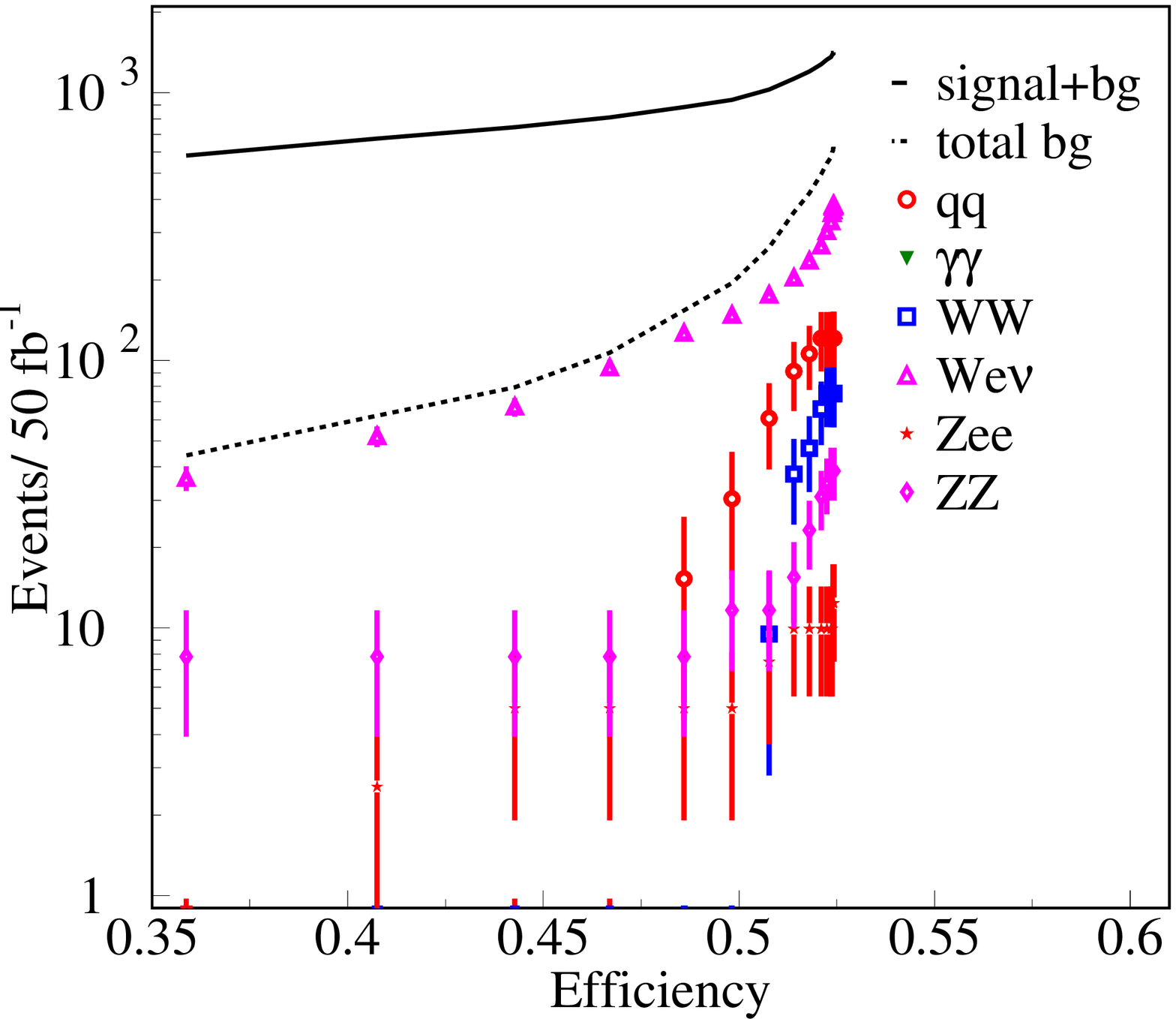}
\end{minipage}
\begin{minipage}{0.49\textwidth}
\vspace*{-1cm}
\includegraphics[width=\textwidth]{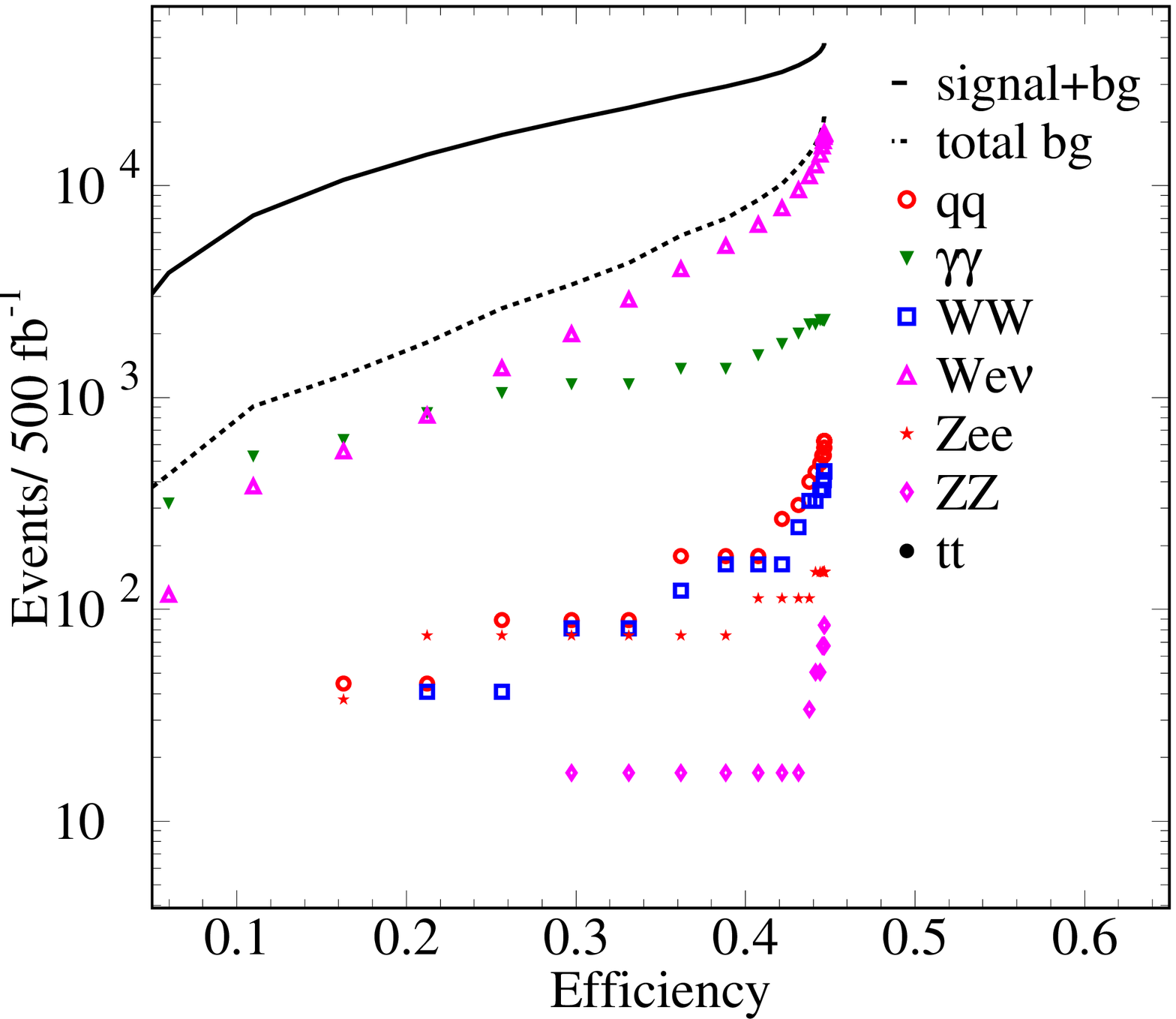}
\end{minipage}
\vspace*{-0.2cm}
\caption{Expected background events as a function of the signal efficiency for 50~fb$^{-1}$ at $\sqrt{s} = 260$ GeV (left plot) 
        and for 500~fb$^{-1}$ at $\sqrt{s} = 500$ GeV (right plot).}
\label{fig:perf}
\end{figure}

\vspace*{-4mm}
The IDA has been applied in two steps in order to optimize the performance, as shown in 
Fig.~\ref{fig:ida} (right plot). 
In the first step, a cut was applied on the IDA\_1 output variable such that 99.5\% 
of the signal events remain.
This leads to 44.3\% signal efficiency and about 17900 background events per 500 fb$^{-1}$.
These remaining signal and background events have been passed to the second IDA step.
A cut on the IDA\_2 output variable (Fig.~\ref{fig:ida2}) determines the final selection efficiency and the 
corresponding expected background. The resulting performance is shown in Fig.~\ref{fig:perf}.

\clearpage

\clearpage

\vspace*{-7mm}
\section{Conclusions}
\vspace*{-4mm}
The IDA leads to a good signal over background ratio
for the investigated Supersymmetric scenario with a scalar top mass of 122.5~GeV and a neutralino
mass of 107.2~GeV (Fig.~\ref{fig:perf}).
We compare the IDA results with a previous analysis using sequential cuts, which for 500~fb$^{-1}$
at $\sqrt{s}=500$ GeV obtained a signal efficiency of 18\% (10620 signal events) with about 6000 background events.
For a similar background level, the IDA achieves a signal efficiency of 36\%,
while for a similar signal efficiency of 18\%, the background is reduced to only about 1400 events with the IDA.
The large sensitivity increase due to the IDA method is similar to the result for 50~fb$^{-1}$
at $\sqrt s = 260$~GeV, where for\,50\%\,signal\,efficiency\,(800\footnote{We improved the previous cross section of 22.5~fb~\cite{lcws06} with the inclusion of QCD corrections to 32~fb.}\,signal\,events)\,about\,200\,background\,events\,(mostly\,$\rm We\nu$) are expected~\cite{lcws06}.
The analysis is a step towards a precise scalar top mass measurement.
The expected uncertainty in the light scalar top mass measurement dominates the 
uncertainty in the dark matter prediction from the co-annihilation process~\cite{carena,sopczak}.
A further study will also include a systematic error analysis.
A much reduced uncertainty on the dark matter prediction is expected~\cite{ayres}.
Further plans are to focus on the vertex detector design, including the implementation of a new 
LCFI~\cite{lcfi} c-quark tagging algorithm.

\vspace*{-4mm}
\begin{theacknowledgments}
\vspace*{-4mm}
AS would like to thank the LCFI colleagues for fruitful discussions 
on the analysis during the preparation of the presentation,
and the organizers of the conference for their hospitality.
\end{theacknowledgments}

\end{document}